# Atomic level structure of Ge-Sb-S glasses: chemical short range order and long Sb-S bonds

I. Pethes[a*], V. Nazabal[b,c], J. Ari[b,d], I. Kaban[e], J. Darpentigny[f], E. Welter[g], O. Gutowski[g], B. Bureau[b], Y. Messaddeq[d], P. Jóvári[a]

[a]Wigner Research Centre for Physics, Hungarian Academy of Sciences, H-1525 Budapest, POB 49, Hungary

[b]Institut Sciences Chimiques de Rennes, UMR-CNRS 6226, Campus de Beaulieu, Université de Rennes 1, 35042 Rennes Cedex, France

[c]Department of Graphic Arts and Photophysics, Faculty of Chemical Technology, University of Pardubice, Studentská 573, 53210 Pardubice, Czech Republic

[d]Center for Optics, Photonics and Lasers (COPL), 2375 rue de la Terrasse, Université Laval, Québec (Qc), Canada

[e]IFW Dresden, Institute for Complex Materials, Helmholtzstr. 20, 01069 Dresden, Germany

[f]Laboratoire Léon Brillouin, CEA-Saclay 91191 Gif sur Yvette Cedex France

[g]Deutsches Elektronen-Synchrotron − A Research Centre of the Helmholtz Association, Notkestraße 85, D-22607 Hamburg, Germany

Abstract

The structure of $Ge_{20}Sb_{10}S_{70}$, $Ge_{23}Sb_{12}S_{65}$ and $Ge_{26}Sb_{13}S_{61}$ glasses was investigated by neutron diffraction (ND), X-ray diffraction (XRD), extended X-ray absorption fine structure (EXAFS) measurements at the Ge and Sb K-edges as well as Raman scattering. For each composition, large scale structural models were obtained by fitting simultaneously diffraction and EXAFS data sets in the framework of the reverse Monte Carlo (RMC) simulation technique. Ge and S atoms have 4 and 2 nearest neighbors, respectively. The structure of these glasses can be described by the chemically ordered network model: Ge-S and Sb-S bonds are always preferred. These two bond types adequately describe the structure of the stoichiometric glass while S-S bonds can also be found in the S-rich composition. Raman scattering data show the presence of Ge-Ge, Ge-Sb and Sb-Sb bonds in the S-deficient glass but only Ge-Sb bonds are needed to fit diffraction and EXAFS datasets. A significant part of the Sb-S pairs has 0.3-0.4 Å longer bond distance than the usually accepted covalent bond length (~2.45 Å). From this observation it was inferred that a part of Sb atoms have more than 3 S neighbors.

Introduction

Chalcogenide glasses have received a broad attention due to their superior optical properties, such as their wide transparency window in the mid-infrared range, high linear and nonlinear refractive indices or unique photosensitivity [1]. Their tunable passive optical properties make them suitable for extensive optical usage, such as IR lenses, optical fibers or waveguides [2] optical fiber amplifiers, sensors or lasers [3].

With the growth of the market in the field of car night vision assistance, the 2020-2025 IR market prospects of Nippon Electric Glass highlight the development of Ge-Sb-S lenses selected for their low toxicity compared to other commercial chalcogenide lenses based on arsenic and selenium. This Ge-Sb-S system also has a significant interest in the field of sensors and nonlinear optics for their extended transparency and their high nonlinear refractive index [4, 5].

The physicochemical properties of glasses are determined by their atomic structure associated with the bonding nature (chemical bonds, structural motifs and their arrangement).

Several structural models of network glasses have been proposed over the decades since the early work of Zachariasen [6]. The most popular among them are the topologically ordered network model [7 - 9] (TONM) and the chemically ordered network model [10] (CONM).

Chalcogenide glasses are essentially covalently connected networks. In glasses consisting of elements from the 14-15-16 groups of the periodic table, (such as Ge-Sb-S glasses) the total coordination number of the elements ($N_i$) mostly follows the Mott-rule [11]: it equals to 8-$N$, where $N$ is the number of the valence electrons of the element (see e.g Ref. [12] and references therein). According to the TONM both homonuclear and heteronuclear bonds are allowed and the physical properties of the glass are determined by the mean coordination number (which is the concentration weighted sum of the coordination numbers of the elements). In contrast, in the CONM there are bond preferences: heteronuclear bonds are favored against homonuclear bonds. As a consequence of the CONM model the glass structure is a network of building blocks (structural units). For an M-X-Ch glass, where M, X and Ch denotes the element from group 14, 15 and 16 respectively, these structural units are typically $MCh_{4/2}$ and $XCh_{3/2}$. According to the CONM model Ch-Ch bonds can be found only in Ch-rich (over-stoichiometric) compositions, while M-M, M-X or X-X bonds are present only in chalcogen deficient glasses.

There are several experimental techniques that can help to determine the structure of a glass, such as X-ray diffraction (see e.g. Ref. [13] and references therein), neutron diffraction (often with isotope substitution, e.g. Ref. [14]), extended X-ray absorption fine structure (EXAFS, e.g. Refs. [15, 16]),

Raman scattering (see e.g. Refs. [17, 18]) or nuclear magnetic resonance (NMR, see e.g. Refs. [19 - 22]). A correct description of a three (or more) component system usually requires utilization of several experimental data sets together. The reverse Monte Carlo simulation technique [23] was applied successfully to generate three dimensional particle configurations compatible simultaneously with data sets obtained by various experimental techniques. This method has already been applied to the study of the structure of Ge-Sb-Te and Ge-Sb-Se glasses (see e.g. Refs. [12, 24]), but there is no similar report about Ge-Sb-S glasses.

The structure of Ge-Sb-S glasses was investigated previously by XRD [25 - 28], ND [29, 30], EXAFS [25, 28] and Raman spectroscopy [31 - 37]. Their results were interpreted mostly in the frame of the CONM: the main building blocks of the network are [$GeS_{4/2}$] and [$SbS_{3/2}$] units as showed e.g. in Refs. [25, 29 - 31, 34 - 36]. S-S bonds were found in S-rich compositions [31, 34]. Ge-Ge, Ge-Sb or Sb-Sb bonds were reported in S-deficient samples [27, 31 - 33] and in stoichiometric compositions as well [25, 34], which shows some chemical disorder.

In this paper we report on a complex structural investigation of Ge-Sb-S glasses. Three samples with approximate composition $Ge_{2x}Sb_xS_{100-3x}$ ($x \approx$ 10, 12, 13) were investigated by ND, XRD and EXAFS at Ge and Sb K-edge as well. The reverse Monte Carlo method was applied to fit the four experimental data sets simultaneously. The obtained atomic configurations were used to determine the short range structural parameters: bonding preferences, bond distances and coordination numbers.

## Experimental

*Glass synthesis and characterization*

The $Ge_{20}Sb_{10}S_{70}$, $Ge_{23}Sb_{12}S_{65}$ and $Ge_{26}Sb_{13}S_{61}$ glasses were prepared by the conventional melt-quench method from a mixture of high-purity elemental germanium, antimony and sulfur purified by distillation. The sealed silica ampoules were slowly heated and homogenized in a rocking furnace for 12 h at high temperature (850°C). Then, the ampoules with melt were quenched into water, and annealed near glass transition temperatures. Investigated compositions and their densities are presented in Table 1.

Table 1. Mass and number densities of the investigated glass compositions.

| Nominal composition | Composition | Mass density (± 0.01 g/cm$^3$) | Number density (atoms/Å$^3$) | $T_g$ (± 2 °C) |
| --- | --- | --- | --- | --- |
| $Ge_{20}Sb_{10}S_{70}$ | $Ge_{19.7}Sb_{9.8}S_{70.5}$ | 3.01 | 0.03715 | 266 |
| $Ge_{23}Sb_{12}S_{65}$ | $Ge_{23.3}Sb_{11.4}S_{65.3}$ | 3.17 | 0.03692 | 342 |
| $Ge_{26}Sb_{13}S_{61}$ | $Ge_{25.8}Sb_{13.2}S_{61}$ | 3.36 | 0.03716 | 313 |

*X-ray diffraction measurements*

X-ray diffraction structure factors were measured at beamline P07 of Petra III (Hamburg, Germany). Samples were ground and filled into quartz capillaries (wall thickness 20 μm, diameter 2 mm). The energy of incident radiation was 82.9 keV (λ=0.1496 Å). Scattered intensities were recorded by a Perkin-Elmer 1621 type area detector. Sample-to-detector distance and detector angle was determined by measuring a $CeO_2$ standard. Raw data were corrected for background scattering. The resulting intensities were integrated using the Fit2D program [38] and corrected for Compton scattering and fluorescence.

*Neutron diffraction measurements*

The neutron diffraction experiment was carried out at the 7C2 liquid and amorphous diffractometer of LLB (Saclay, France). The wavelength of neutrons was 0.724Å. Powdered samples were filled into vanadium sample holders (wall thickness 0.1 mm, diameter 5mm). Raw data were corrected for detector efficiency, background and multiple scattering.

*EXAFS measurements*

Ge K-edge EXAFS spectra were measured in fluorescence mode at beamline P65 of the Petra III source. Powdered samples were mixed with cellulose and pressed into tablets. The absorption of the pellets was around above the selected absorption edge. Monochromatic radiation was obtained by a Si111 double crystal monochromator. The Sb K-edge EXAFS data were recorded at the same beamline in transmission mode, using a Si311 double crystal monochromator.

$\chi(k)$ curves were obtained using the Viper program [39]. Raw $\chi(k)$ signals were first forward Fourier-transformed using a Kaiser-Bessel window. The resulting r-space curves were back transformed using a rectangular window over 1.1-2.4 Å and 1.1-3.3 Å for Ge and Sb-edge data, respectively.

*Raman scattering*

Raman spectra were collected using LabRam HR800 (Horiba Jobin-Yvon) spectrometer at room temperature using 785 nm laser excitation. The power of the laser was reduced with optical density filters to avoid possible photoinduced changes.

Reverse Monte Carlo simulations

The reverse Monte Carlo simulation technique [23] is an inverse method to get large three dimensional atomic configurations consistent with experimental (or theoretical) data. It can be used in principle with any quantity that can be expressed directly from the atomic coordinates, such as structure factors from diffraction (ND or XRD) measurements or EXAFS data sets. Over the past three decades the RMC method has been successfully applied to a wide collection of systems, from molecular liquids (e.g. Refs. [40, 41]), through metallic and covalent glasses (e.g. Refs. [42, 43]) to disordered crystalline materials (e.g. Refs. [44, 45]).

In the RMC modeling the partial pair correlation functions (PPCF) ($g_{ij}(r)$) are calculated from the generated particle coordinates. The partial structure factors ($S_{ij}(Q)$) are related to the PPCFs by equation (1):

$$S_{ij}(Q) - 1 = \frac{4\pi\rho_0}{Q}\int_0^\infty r\big(g_{ij}(r) - 1\big)\sin(Qr)\,dr. \tag{1}$$

Here $Q$ is the modulus of the scattering vector and $\rho_0$ is the average number density. The XRD and ND total structure factors can be expressed with the partial structure factors $S_{ij}(Q)$ as:

$$S(Q) = \sum_{i \leq j} w_{ij}^{X,N}(Q) S_{ij}(Q). \tag{2}$$

$w_{ij}^{X,N}$ denotes the X-ray and neutron scattering weights. For X-rays it is given by equation (3):

$$w_{ij}^X(Q) = (2 - \delta_{ij})\frac{c_i c_j f_i(Q) f_j(Q)}{\sum_{ij} c_i c_j f_i(Q) f_j(Q)}. \tag{3}$$

Here $\delta_{ij}$ is the Kronecker delta, $c_i$ denotes the atomic concentration, $f_i(Q)$ is the atomic form factor. The neutron weight factors are:

$$w_{ij}^N = (2 - \delta_{ij})\frac{c_i c_j b_i b_j}{\sum_{ij} c_i c_j b_i b_j}, \tag{4}$$

where $b_i$ is the coherent neutron scattering length.

The experimental X-ray absorption coefficient is converted into the EXAFS signal $\chi_i(k)$ as a function of the wavenumber $k$ of the photoelectron. $\chi_i(k)$ is related to the PPCFs by [46]:

$$\chi_i(k) = \sum_j 4\pi\rho_0 c_j \int_0^R r^2 \gamma_{ij}(k,r) g_{ij}(r)\,dr \tag{5}$$

Here $i$ refers to the absorber atom, $\gamma_{ij}(k,r)$ is the photoelectron backscattering matrix, which gives the $k$-space contribution of a $j$-type backscatterer at distance $r$ from the absorber atom. Elements of $\gamma_{ij}(k,r)$ matrix were calculated for each $i$-$j$ pair by the FEFF8.4 program [47].

During the simulations particles are moved around randomly in the simulation box to minimize the differences between the experimental ($S_{exp}(Q)$, $\chi_{exp}(k)$) and RMC model ($S_{mod}(Q)$, $\chi_{mod}(k)$) curves. The undeniable advantage of this method is that the experimental data sets are fitted simultaneously and the obtained final particle configurations are compatible with all of the input data (within their experimental uncertainties).

The quality of the fits may be compared via their 'goodness-of-fit' ($R$-factor) values:

$$R = \frac{\sqrt{\sum_i \left(S_{mod}(Q_i) - S_{exp}(Q_i)\right)^2}}{\sqrt{\sum_i S_{exp}^2(Q_i)}} \qquad (6)$$

where $Q_i$ denote the experimental points.

Further physical and chemical properties, such as density, preferred bond angles, or coordination numbers can be taken into account. From the final configurations structural characteristics (partial pair correlation functions, nearest neighbor distances, average coordination numbers etc.) can be calculated. The Ge and Sb EXAFS data sets and the two (XRD and ND) structure factors for each composition were fitted by the RMC++ code [48]. The simulation boxes contained 10000 atoms for the test runs and 25000 atoms for the final runs presented here. Initial configurations were obtained by placing the atoms randomly into the simulation box and moving them around to fulfill the minimum interatomic distance (cutoff) requirements. The cutoff distances are shown in Table 2.

Table 2. Minimum interatomic distances (cut-offs) applied in simulation runs (in Å)

| Pair | Ge-Ge | Ge-Sb | Ge-S | Sb-Sb | Sb-S | S-S |
|---|---|---|---|---|---|---|
| Bond is allowed | 2.25 | 2.35 | 2.0 | 2.7 | 2.25 | 1.95 |
| Bond is forbidden | 2.75 | 2.9 | - | 3.15 | - | 2.9 |

Dedicated simulation runs were carried out to determine which bond types are required to get reasonable fits. A bond type was excluded by using cutoff values higher than the expected bond distance. The Ge-S and Sb-S bonds were always allowed in accordance with the CONM, where these bonds are preferred. In the final models Ge-Ge and Sb-Sb bonds were forbidden in all compositions. Ge-Sb bonds were allowed only for the S-deficient sample. The S-S bonds were forbidden in the S-deficient and the stoichiometric composition, but were necessary in the S-rich glass. An additional constraint was used to avoid the relatively high uncertainty of $g_{SS}$ (which originates from the small number of S-S pairs, and the moderate weights and presence of $g_{SS}$ in the measurements): the S-S bond length was constrained to be around 2.05 Å [49].

Some additional coordination constraints were always applied to avoid unrealistically low coordination numbers (2 or less for Ge, 1 or 0 for Sb and 0 for S).

The number of accepted moves was typically around 1-2× $10^7$. The $\sigma$ parameters, which influence the tightness-of-fit [23, 50], were reduced gradually during the simulation runs, resulting in a progressively improving fit to the target curves. Final $\sigma$ values were 1-2 × $10^{-3}$ for the diffraction data sets and 1-2 × $10^{-5}$ for the EXAFS data sets.

The average coordination numbers ($N_{ij}$) were calculated from the PPCFs by integrating them up to the minimum between the first and second coordination sphere. The uncertainties of the average coordination numbers were estimated by dedicated simulation runs in which coordination constraint was used for the value of the investigated average coordination number to force it to be higher or lower.

The different test models were classified according to their $R$-factors which were compared to the $R$-factors of the unconstrained model.

Results and discussion

*Raman scattering spectroscopy analyses*

The Raman spectra of the investigated glasses are shown in Fig. 1. For these germanium based glasses, three main vibration bands are present at 338, 369 and 405-430 cm$^{-1}$ in agreement with GeS$_2$ glass [10]. The strongest at 338 cm$^{-1}$ can be associated with the $\nu_1(A_1)$ symmetric stretching mode of [GeS$_{4/2}$] tetrahedron. The shoulder at 369 cm$^{-1}$ can be associated to the $\nu_{c1}(A_{c1})$ "companion" mode of the $\nu_1$ mode, often linked to the vibrations of tetrahedra [GeS$_{4/2}$] bound by their edges [51]. The broad band located between 405-435 cm$^{-1}$ is related to the stretching vibration of S$_3$Ge-S-GeS$_3$ structural units where tetrahedra are connected by their corners and other vibration modes of [GeS$_{4/2}$] units. The antimony presence in germanium based glasses induces the emergence of a vibration band around 302 cm$^{-1}$. This band has previously been associated with the vibration of the [SbS$_3$] trigonal pyramids [31, 52]. Because of the over-stoichiometry in sulfur of the Ge$_{20}$Sb$_{10}$S$_{70}$ glass, the vibration mode of the S-S bonds is observable at around 473 cm$^{-1}$ which can be associated with S-S dimers or small chains connecting two tetrahedral or trigonal units and to the presence of S$_8$ rings whose main vibration bands are centered at around 151, 218 and 473 cm$^{-1}$ [53].

In case of the S-poor Ge$_{26}$Sb$_{13}$S$_{61}$ glass, the Raman spectrum shape in the low frequency range (150-

260 cm$^{-1}$) is linked to the formation of homonuclear bonds [33, 54]. Firstly, the Raman band at 254 cm$^{-1}$ is the signature of Ge-Ge bonds in structural entities containing fewer than four S atoms around Ge of Ge(Sb)$_x$S$_{(3-x)}$-Ge-GeS$_{(3x)}$Ge(Sb)$_x$ type [10]. The band at 208 cm$^{-1}$ could be related to the vibration of bonds linking two dissimilar elements like Sb-Ge [54]. Finally, the Raman band at 163 cm$^{-1}$ can be connected with vibrations of Sb-Sb bonds present in films prepared by pulsed laser deposition [55].

*RMC investigations*

The measured total structure factors ($S(Q)$), and $k^3$–weighted, filtered EXAFS curves ($k^3\chi(k)$) of the investigated glasses are plotted in Figures 2 and 3. Model configurations were obtained by simultaneous fitting of the four experimental data sets for each composition. Various models were tested to determine the necessary bond types. Ge-S and Sb-S bonds are found to be required in all compositions. In addition, S-S bonding in Ge$_{20}$Sb$_{10}$S$_{70}$ and Ge-Sb bonding in Ge$_{26}$Sb$_{13}$S$_{61}$ are also needed to get adequate fits. The simulated curves obtained by applying the most appropriate models are presented in Figures 2 and 3.

The observed bond type requirements are in total agreement with the predictions of the CONM. According to the Mott-rule the Ge$_{23}$Sb$_{12}$S$_{65}$ glass is nearly stoichiometric; thus the structure of this glass can be described by GeS$_4$ and SbS$_3$ building blocks, which share their S atoms. The number of S-S pairs is undetectably small, similarly to the numbers of Ge-Ge, Ge-Sb or Sb-Sb pairs.

The S excess in an S-rich glass results in the appearance of S-S bonds, as it is found in the Ge$_{20}$Sb$_{10}$S$_{70}$ glass. The lack of Ge-Ge, Ge-Sb and Sb-Sb pairs in the structure of this glass is also consistent with the CONM.

The Ge$_{26}$Sb$_{13}$S$_{61}$ glass is chalcogen poor, according to the Mott-rule. Thus, beside Ge-S and Sb-S bonds some Ge-Ge, Ge-Sb or Sb-Sb bonds must be formed to fulfill the bond requirements of the elements. S-S bonds are not expected in this composition in the CONM. Indeed, simulations showed that S-S bonds are not required to get adequate fits. The necessity of Ge-Ge, Ge-Sb and Sb-Sb bonds was tested one by one. It was found that the quality of the fits is significantly better when Ge-Sb pairs are present, while allowing Ge-Ge or Sb-Sb pairs does not reduce the *R*-factors of the data sets. Ge-Sb bonds were also found in S-deficient Ge-Sb-S glasses earlier [27].

Partial pair correlation functions are shown in Figure 4. The $g_{ij}(r)$ curves for the allowed bond types have sharp peaks in the 2.0 Å ≤ $r$ ≤ 3.0 Å region. Positions of the first peaks (the average nearest neighbor distances) are collected in Table 3.

Table 3. Nearest neighbor distances (in Å) in the studied Ge-Sb-S glasses. (The S-S bond lengths are constrained!) The uncertainty of distances is usually ± 0.02 Å.

| Pair | Glass composition | | |
|---|---|---|---|
| | $Ge_{20}Sb_{10}S_{70}$ | $Ge_{23}Sb_{12}S_{65}$ | $Ge_{26}Sb_{13}S_{61}$ |
| Ge-Sb | - | - | 2.61 |
| Ge-S | 2.23 | 2.22 | 2.23 |
| Sb-S | 2.45 | 2.45 | 2.45 |
| S-S | 2.05 | - | - |

The first coordination shell of Ge atoms contains S atoms in $Ge_{20}Sb_{10}S_{70}$ and $Ge_{23}Sb_{12}S_{65}$, while in $Ge_{26}Sb_{13}S_{61}$ glass S and Sb atoms can be found around Ge atoms. The Ge-S and Ge-Sb coordination numbers were obtained by counting particles up to 2.9 Å and are listed in Table 4. The total coordination number of Ge ($N_{Ge} = N_{GeSb} + N_{GeS}$) is 3.8, 3.96 and 3.96 for $Ge_{20}Sb_{10}S_{70}$, $Ge_{23}Sb_{12}S_{65}$ and $Ge_{26}Sb_{13}S_{61}$, respectively. These values, which were obtained *without any coordination constraints*, are very close to 4, the value predicted by the Mott-rule.

Table 4. Coordination numbers of the investigated glasses obtained by RMC simulation without any coordination constraints. The allowed bonds were Ge-S, Sb-S, S-S for the $Ge_{20}Sb_{10}S_{70}$ glass, Ge-S, Sb-S for the $Ge_{23}Sb_{12}S_{65}$ glass and Ge-Sb, Ge-S, Sb-S for the $Ge_{26}Sb_{13}S_{61}$ composition. Uncertainties of the coordination numbers are shown in parentheses. (For the Sb-S partials the coordination numbers are calculated without and with 'long bonds' as well.)

| | Glass composition | | |
|---|---|---|---|
| | $Ge_{20}Sb_{10}S_{70}$ | $Ge_{23}Sb_{12}S_{65}$ | $Ge_{26}Sb_{13}S_{61}$ |
| $N_{GeGe}$ | 0 | 0 | 0 |
| $N_{GeSb}$ | 0 | 0 | 0.29 (±0.15) |
| $N_{GeS}$ | 3.8 (-0.2 +0.6) | 3.96 (±0.1) | 3.67 (-0.1+0.2) |
| $N_{SbGe}$ | 0 | 0 | 0.57 (±0.25) |
| $N_{SbSb}$ | 0 | 0 | 0 |
| $N_{SbS}$ | 2.62/3.26 (-0.45+0.3) | 2.64/3.16 (-0.25+0.4) | 1.7/2.87 (-0.8 +0.2) |
| $N_{SGe}$ | 1.06 (-0.05+0.15) | 1.41 (±0.05) | 1.55 (-0.05+0.1) |
| $N_{SSb}$ | 0.36/0.45 (±0.05) | 0.46/0.55 (±0.05) | 0.37/0.62 (-0.15+0.05) |
| $N_{SS}$ | 0.58 (+0.2-0.4) | 0 | 0 |
| $N_{Ge}$ | 3.8 | 3.96 | 3.96 |
| $N_{Sb}$ | 2.62/3.26 | 2.64/3.16 | 2.27/3.44 |
| $N_{S}$ | 2.00/2.09 | 1.87/1.96 | 1.92/2.17 |

The first coordination shell of Ge atoms does not contain Ge atoms, but there are well-defined Ge-Ge distances in the second coordination shell. The $g_{GeGe}(r)$ curves have a pronounced peak around 3.5 Å. A smaller peak at 2.8 Å is also present, which is most significant in the stoichiometric $Ge_{23}Sb_{12}S_{65}$ composition. By analyzing the configurations, the shorter distance can be identified as the distance of Ge-Ge pairs in edge-shared GeS$_4$ tetrahedra (see Figure 5). The peak around 3.5 Å contains Ge-Ge distances of corner- and edge-shared GeS$_4$ tetrahedra as well. The ratio of Ge atoms participating in edge-shared units is around 42 – 48 %, in the compositions investigated.

The environment of Sb atoms contains only S atoms in Ge$_{20}$Sb$_{10}$S$_{70}$ and Ge$_{23}$Sb$_{12}$S$_{65}$, while in Ge$_{26}$Sb$_{13}$S$_{61}$ glass S and Ge atoms can be found around Sb atoms. The total coordination number of Sb ($N_{Sb} = N_{SbGe} + N_{SbS}$) – counting Sb-S pairs up to 3.1 Å – is 3.26, 3.16 and 3.44 (in the same order as before, see Table 4). The values of the S-rich and stoichiometric glasses are close to the value predicted by the Mott-rule (3), the difference is higher for the S-deficient glass.

Sb-Sb pairs sharing two common S atoms (edge sharing SbS$_x$ units) can be found in all glasses. The ratio of Sb atoms participating in these units is in the 5% - 11% range.

The S atoms are connected only to Ge and Sb atoms in Ge$_{23}$Sb$_{12}$S$_{65}$ and Ge$_{26}$Sb$_{13}$S$_{61}$, while in Ge$_{20}$Sb$_{10}$S$_{70}$ glass S-S pairs are also present. The S-S pairs were counted up to 2.2 Å to calculate the $N_{SS}$ coordination number. The total coordination number of S atoms ($N_S = N_{SGe} + N_{SSb} + N_{SS}$) is 2.09, 1.96 and 2.17 for the three glasses (in the same order as before). Overall, it can be concluded that in ternary Ge-Sb-S glasses the atoms mostly obey the Mott-rule.

The Ge-S bond distance (2.22-2.23 Å, see Table 3) is similar to those found earlier in binary Ge-S systems (2.21-2.23 Å e.g. by ND [56], by ND and high energy X-ray diffraction measurements [49] or by pulsed ND [57]). In ternary Ge-Sb-S glasses slightly longer values were estimated from single measurements earlier: 2.255 Å by EXAFS [25], 2.26 Å by XRD [27], 2.24 Å by ND [30], and 2.24 Å by EXAFS [28].

The Ge-Sb bond distance is 2.61 Å. Earlier in Ge-Sb-S glasses Ge-Sb bond length were found 2.765 Å by EXAFS [25], 2.65 Å by XRD [27]. 2.64 Å Ge-Sb bond distance was reported for Ge-Sb-Se glasses by RMC method with EXAFS, ND and XRD measurements [12].

The first peak of the Sb-S PPCF is around 2.45 Å, which is lower than that found in Ge-Sb-S by ND (2.48 Å) [30], but similar to the value observed by EXAFS for Sb-S systems (2.45-2.46 Å) [58].

The shape of the $g_{SbS}(r)$ PPCF is somewhat surprising: behind the main peak between 2.3 Å and 2.6 Å there is a second peak between 2.6 Å and 3.1 Å. The $N_{SbS}$ average coordination numbers for the main peak are 2.62, 2.64 and 1.7 for Ge$_{20}$Sb$_{10}$S$_{70}$, Ge$_{23}$Sb$_{12}$S$_{65}$ and Ge$_{26}$Sb$_{13}$S$_{61}$, respectively. For the second region these values are (in the same sequence) 0.64, 0.52 and 1.17. It often happens that small peaks or shoulders are present after the well-defined main peak in the PPCFs of RMC-generated models of covalent glasses (see e.g the $g_{GeS}(r)$ curves in Figure 4). However, the ratios of the coordination numbers corresponding to long and normal bonds are usually significantly lower, than the values of $N_{SbS-long}$ / $N_{SbS-normal}$ (which are 0.24, 0.20 and 0.69).

Dedicated simulation runs were performed to decide whether the second peak in the $g_{SbS}(r)$ curves is a simulation artifact or a real feature of the glasses investigated. In one series of these runs a coordination constraint was applied to eliminate the second peak: it was required that $g_{SbS}(r) = 0$ between 2.65 Å and 3.05 Å. The $R$-factors of the Sb EXAFS data sets increased by 30-100 % for the three compositions.

The position of the second peak is around 2.80-2.85 Å. This value is close to the expected distance of possible Sb-Sb pairs. In another series of test runs long Sb-S bonds were forbidden, but Sb-Sb pairs were allowed, moreover the Sb-Sb coordination number was constrained to be significant (from 0.2 up to 1.0). The quality of the fits did not improve, the $R$-factor of the XRD data set increased significantly. In a third test run the Sb EXAFS data set was omitted and only the two structure factors (ND and XRD) and the Ge EXAFS data sets were fitted together. In this case in the 2.6 Å $\leq r \leq$ 3.0 Å region a broad shoulder appeared in the $g_{SbS}(r)$ curve, as it is shown in Figure 6.

The presence of the long Sb-S bonds can be interpreted considering the structure of crystalline $Sb_2S_3$. The structure of crystalline $Sb_2S_3$ contains parallel $(Sb_4S_6)_n$ chains, which are linked to form crumpled sheets. Sb atoms are distributed over two different crystallographic sites [59 - 61]. One shows a trigonal $SbS_3$ pyramid with the Sb atom at the vertex. This pyramid is slightly distorted; the interatomic Sb-S distances are 2.52, 2.54 and 2.54 Å. The other part of the Sb atoms is in the form of $SbS_5$ distorted square pyramids (with the Sb atom slightly below the base center). The interatomic Sb-S distances in the $SbS_5$ pyramids are 2.46, 2.68, 2.68, 2.85, 2.85 Å.

The broad distribution of Sb-S bond distances in glassy Ge-Sb-S can originate from the combined presence of $SbS_3$ and $SbS_5$ units. The higher coordination number of Sb atoms suggests the same. The combination of $SbS_3$ and $SbS_5$ (and $SbS_4$) units was observed (by X-ray absorption spectroscopy) earlier in glassy $Sb_2S_3$-$As_2S_3$-$Tl_2S$ systems as well [62].

## Conclusions

The structure of three Ge-Sb-S glasses (an S-poor, an S-rich and a stoichiometric) was investigated by Raman scattering, ND, XRD and EXAFS measurements at the Ge and Sb K-edges. Diffraction and EXAFS data sets were fitted simultaneously by RMC simulation method for each composition. It was shown that constituents mainly obey the Mott-rule but a part of Sb atoms may have more than 3 S neighbors The structure of these glasses can be described by the CONM: Ge-S and Sb-S bonds are the most preferred; S-S bonds are not needed in the S-poor and the stoichiometric glass. Raman scattering data show the presence of Ge-Ge, Ge-Sb and Sb-Sb bonds in the S-deficient composition but only Ge-Sb bonds are needed to fit diffraction and EXAFS datasets. Sb-Sb and Ge-Ge bonds can be eliminated suggesting that the corresponding coordination numbers are below the sensitivity of diffraction and EXAFS data (~0.3). Long Sb-S bonds were observed in all Ge-Sb-S glasses. It is proposed that the presence of these bonds is connected with $SbS_5$ units that may exist in this system similarly e.g. to crystalline $Sb_2S_3$.

## Acknowledgments

The authors thank Czech Science Foundation (GACR, project No. 16-17921S) for financial support of this work. The neutron diffraction experiment was carried out at the ORPHÉE reactor, Laboratoire Léon Brillouin, CEA-Saclay, France. I. P. and P. J. were supported by NKFIH (National Research, Development and Innovation Office) Grant No. SNN 116198.

Figures

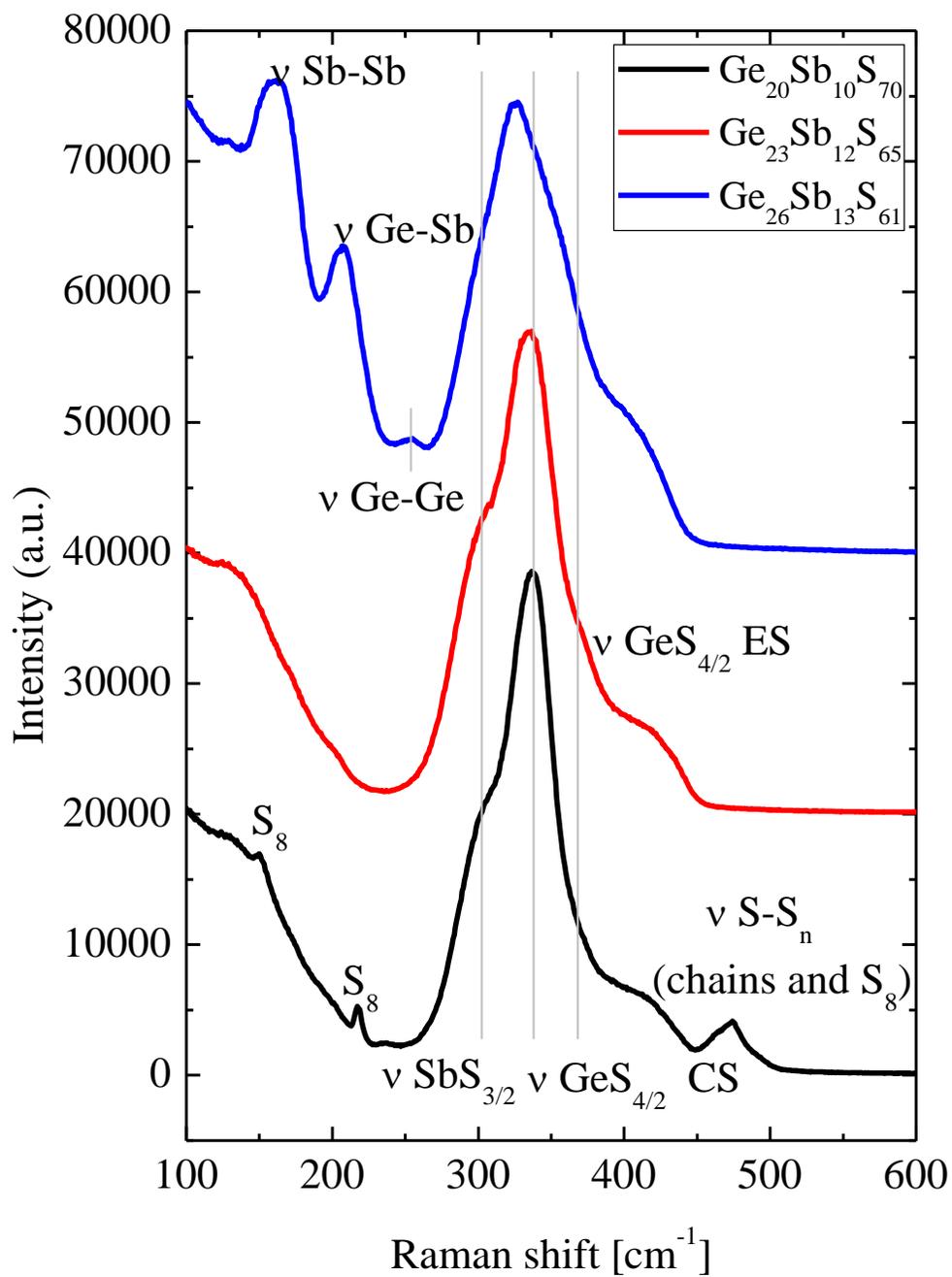

Figure 1. Raman spectra of $Ge_{20}Sb_{10}S_{70}$, $Ge_{23}Sb_{12}S_{65}$, and $Ge_{26}Sb_{13}S_{61}$ glasses.

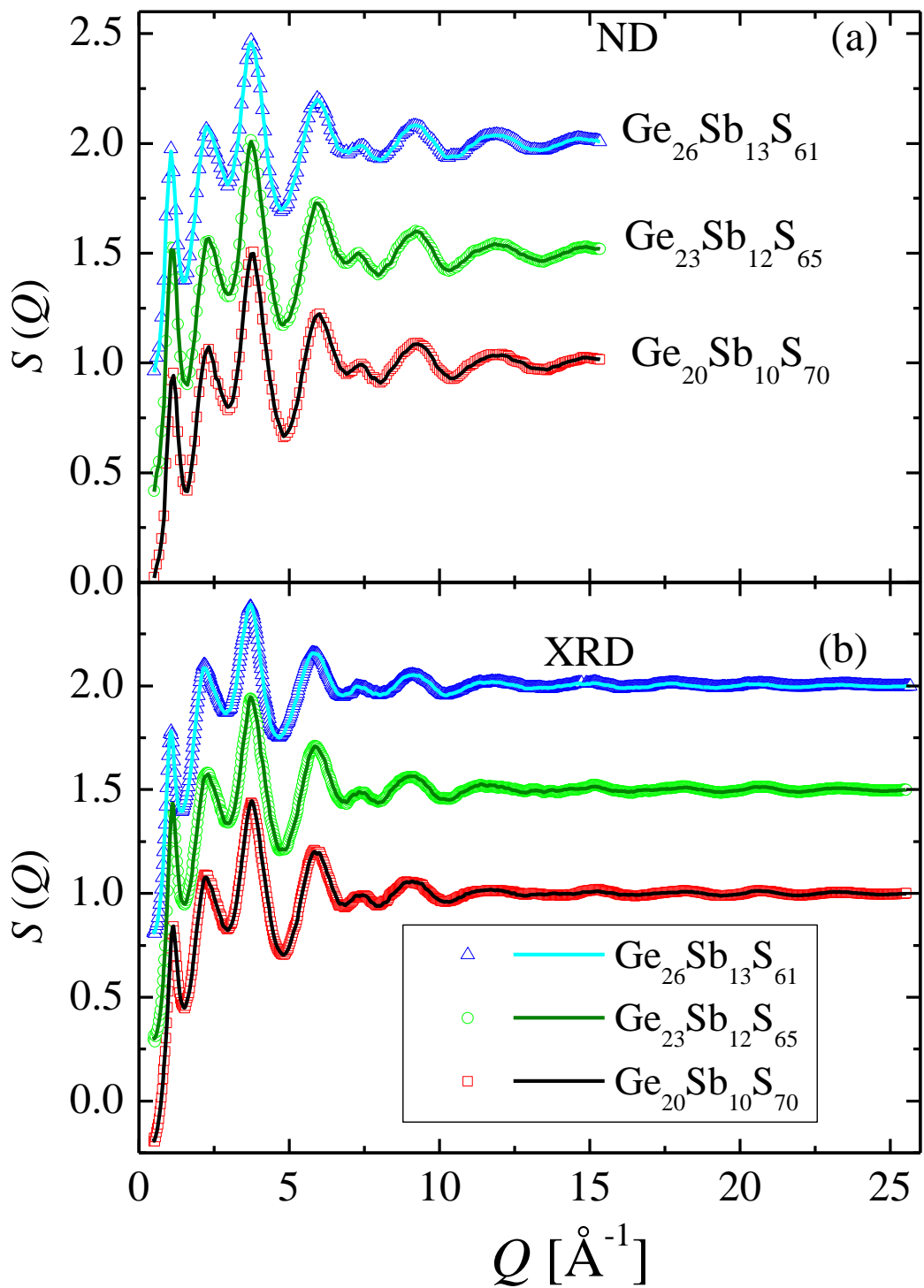

Figure 2. (a) ND and (b) XRD structure factors of the investigated glasses (symbols) and fits (lines) obtained by RMC simulations. (The curves from $Ge_{20}Sb_{10}S_{70}$ to $Ge_{26}Sb_{13}S_{61}$ are shifted upward to improve clarity.)

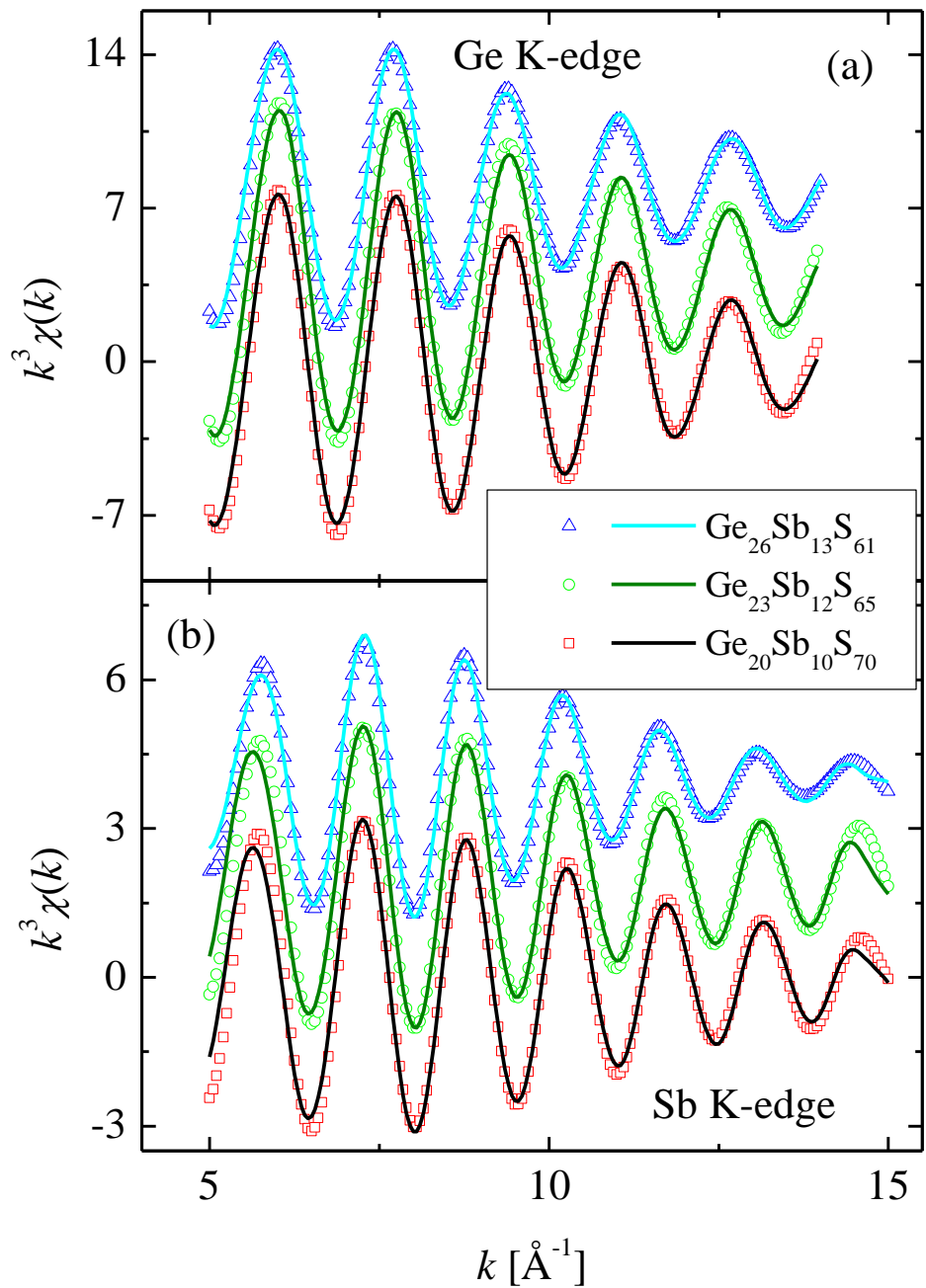

Figure 3. $k^3$–weighted, filtered (a) Ge and (b) Sb K-edge EXAFS spectra of the investigated glasses (symbols) and fits (lines) obtained by RMC simulations. (The curves from $Ge_{20}Sb_{10}S_{70}$ to $Ge_{26}Sb_{13}S_{61}$ are shifted upward to improve clarity.)

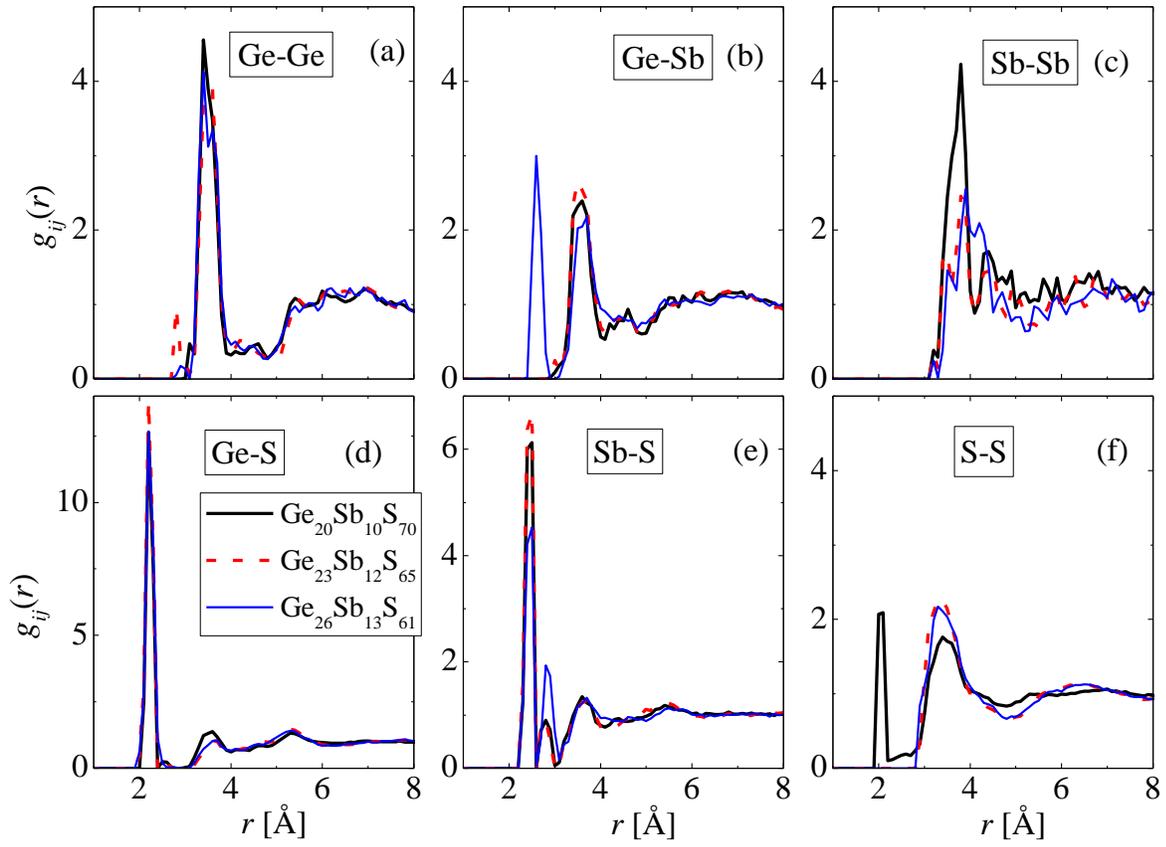

Figure 4. Partial pair correlation functions of the investigated Ge-Sb-S glasses.

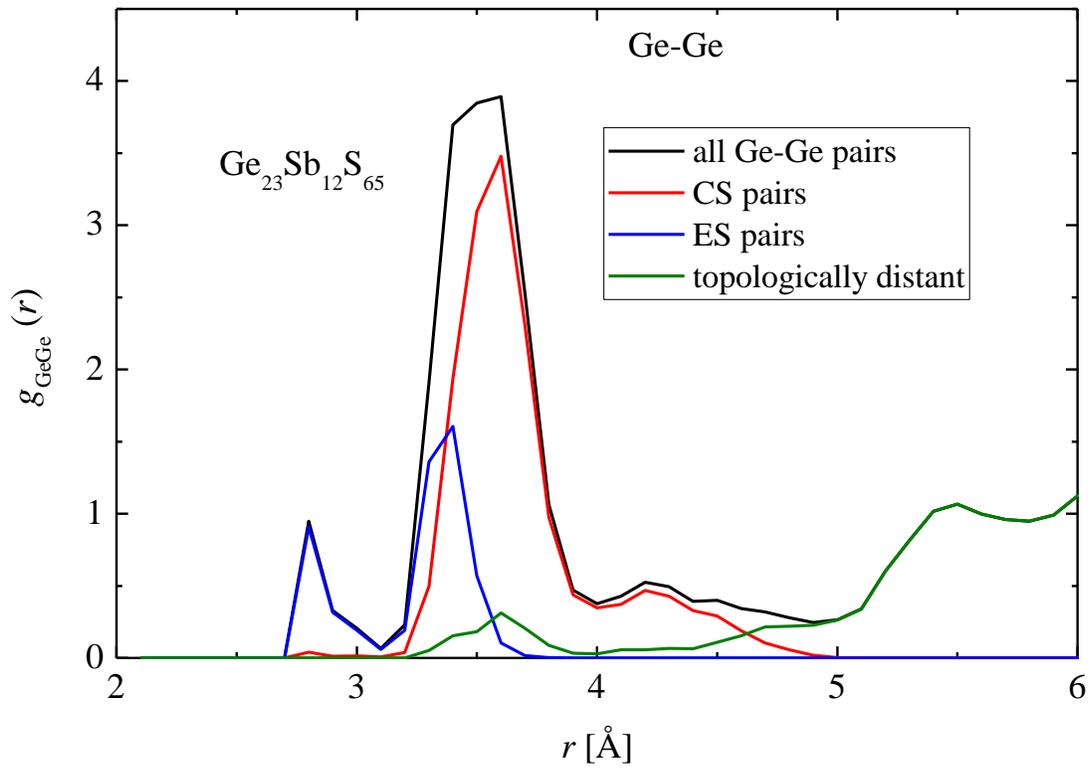

Figure 5. Decomposition of $g_{GeGe}(r)$ of $Ge_{23}Sb_{12}S_{65}$ to contributions from corner sharing (CS) tetrahedra, edge sharing (ES) tetrahedra and topologically distant Ge-Ge pairs.

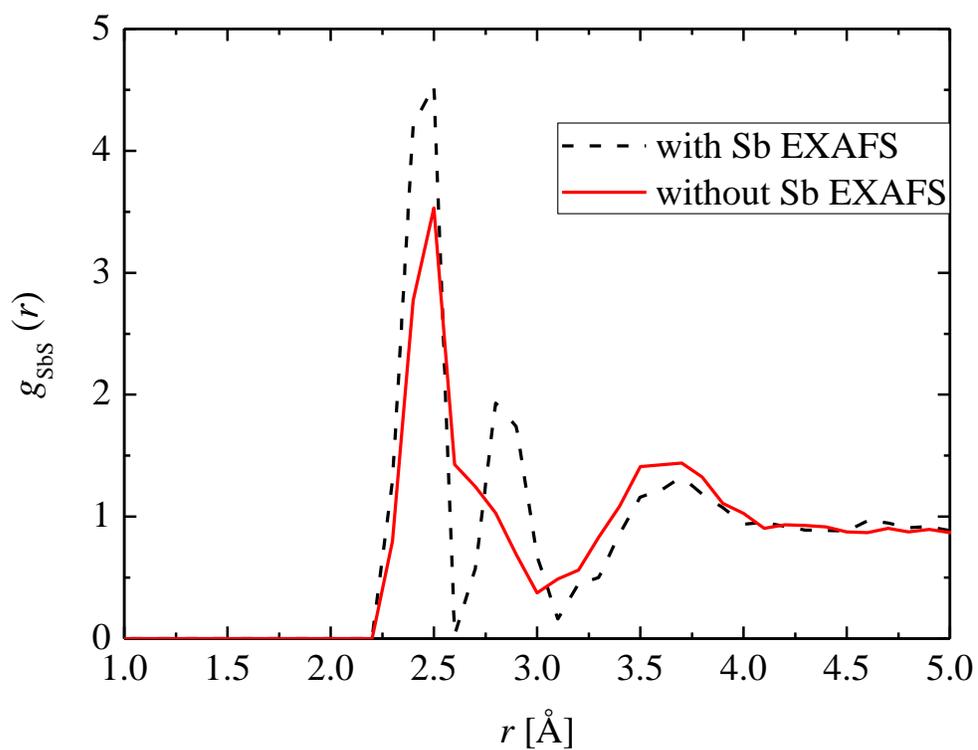

Figure 6. Sb-S partial pair correlation functions of $Ge_{26}Sb_{13}S_{61}$ glass obtained by fitting all four (ND, XRD, Ge and Sb EXAFS) experimental data sets simultaneously (dashed line) and the same PPCF obtained by fitting only ND, XRD and Ge EXAFS data sets (solid line).